
\magnification=1200
\overfullrule=0pt

\font\twelvebf=cmbx12
\font\ibf=cmbxti10

\def\Ac{{\cal A}}
\def\Lc{{\cal L}}
\def\Fc{{\cal F}}
\def\Hc{{\cal H}}

\newfam\bbfam

\def\un{{\rm 1\mkern-4mu I}}

\def\buildrel#1\over#2{\mathrel{
\mathop{\kern 0pt#2}\limits^{#1}}}

\def\build#1_#2^#3{\mathrel{
\mathop{\kern 0pt#1}\limits_{#2}^{#3}}}

\catcode`\@=11
\def\displaylinesno #1{\displ@y\halign{
\hbox to\displaywidth{$\@lign\hfil\displaystyle##\hfil$}&
\llap{$##$}\crcr#1\crcr}}

\baselineskip=14pt

\rightline{CERN-TH/95-274}
\rightline{IHES/P/95/91~~~~}
\rightline{UWThPh-33-1995}
\rightline{hep-th/9510177}
\vskip 0.5cm
\centerline{\twelvebf Simple Field Theoretical Models on}

\smallskip

\centerline{\twelvebf Noncommutative Manifolds}

\bigskip

\centerline{\bf H. Grosse\footnote{${}^*$}{\sevenrm Part of project
No. P8916-PHY of the ``Fonds zur F\"orderung der wissenschaftlichen
Forschung in \"Oster\-reich''.}}

\smallskip

\centerline{Institute for Theoretical Physics, University of Vienna,}

\centerline{Boltzmanngasse 5, A-1090 Vienna, Austria}

\medskip

\centerline{\bf C. Klim\v c\'{\i}k}

\smallskip

\centerline{Theory Division CERN, CH-1211 Geneva 23,}

\centerline{Switzerland}

\smallskip

\centerline{\it and}

\smallskip

\centerline{\bf P. Pre\v snajder}

\smallskip

\centerline{Dept. of Theoretical Physics, Comenius University,}

\centerline{Mlynsk\'a dolina F1, SK-84215 Bratislava, Slovakia}

\bigskip

\centerline{\bf Abstract}

\medskip

We review recent progress in formulating two-dimensional models over
noncommutative manifolds where the space-time coordinates enter in the
formalism as non-commuting matrices. We describe the Fuzzy sphere and a way to
approximate topological nontrivial configurations using matrix
models. We obtain an ultraviolet cut off procedure, which respects
the symmetries of the model. The treatment of spinors results from a
supersymmetric formulation; our cut off procedure preserves even the
supersymmetry.

\vskip 0.5cm
\centerline{\it Based on lectures presented by H. Grosse at Workshops on}
\centerline{\it   Lie Theory and Its Applications in Physics,
Clausthal, Germany, August 1995}
\centerline{\it and on}
\centerline{\it  New trends in Quantum Field Theory, Razlog,
Bulgaria,
August 1995}
\medskip
\leftline{CERN-TH/95-274}
\leftline{October 1995}
\vfill\eject

\noindent {\bf 1. Introduction}

\medskip

We intend to use part of the ideas of noncommutative geometry [1], and
apply them to simple models of quantum field theory. Simple refers
here just to two dimensions. Moreover we shall treat the euclidean
version of the models and compactify space time. Thus we arrive at the
two-sphere $S^2$, although other manifolds can be treated.

\smallskip

Following old ideas of von Neumann, we encode as much structure as
possible, within the commutative algebra of smooth functions defined
over the manifold $M$, which we denote by $\Ac_{\infty}$. Next we shall
approximate this algebra by a sequence of noncommutative algebras
$\Ac_n$, which converge in a specific manner to $\Ac_{\infty}$ for
$n\rightarrow \infty$.

\smallskip

One formulates next the differential geometry on the manifold $M$ in
terms of ope\-rations on $\Ac_{\infty}$. For example vector fields on
$M$ can be identified as derivations on $\Ac_{\infty}$. Differential
forms can be defined as duals of vector fields or with the help of a
Dirac operator plus some additional structure. The last formulation
can be applied in the noncommutative case too, and a differential
calculus over noncommutative algebras results.

\smallskip

The geometry so obtained, is often referred to a pointless geometry.
In the commutative case, abelian ideals of the algebra correspond to
points of the manifold. Our algebras will have no nontrivial ideals.

\smallskip

Next, fields have to be defined as sections of line, respectively,
spinor bundles, or gauge field potentials as connections of principal
bundles. They form modules over $\Ac_{\infty}$ and the natural
noncommutative generalization consists in studying finitely generated
projective modules, which will enter our discussion later on. Finally
an integral calculus is needed, in order to integrate our forms. We
will have no problems in that respect, since all our algebras will be
finite dimensional. Also de Rham cohomology has been generalized to
cyclic cohomology, and cyclic cocycles replace the notion of an
integral.

\smallskip

Let us mention, for later use, a simple example: the differential
calculus on matrix algebras $M_n = {\rm Mat} \, (n, {\bf C})$ [2]. Let the
algebra be generated by $\{ \un ,\lambda_i \}$, $i=1,\ldots n^2 -1$.
Vector fields can be identified as derivations $e_i ={\rm ad} \,
\lambda_i$. A difference to the calculus on manifolds shows up:
derivations do not form anymore a left module.

\smallskip

Differential forms may be introduced by duality:
$$
(d \, \lambda_j) \, (e_i) = e_i \, (\lambda_j) = C_{ijk} \, \lambda_k
\, , \eqno (1) $$
where $C_{ijk}$ denote the structure constants.

\smallskip

Zero-forms are identified with algebra elements $\Omega^0 =M_n$;
one-forms have been just defined $\Omega^1 = \{ a \, db \mid a,b\in M_n
\}$, and there is no difficulty in extending $d$, such that the full
differential complex results. Knowing one-forms allows to define a
covariant derivative $D=d+A$ and a curvature two-form $F=D^2 \equiv
dA+A^2$. Gauge transformations can be defined and an action
functional for gauge fields can be written down. Thus one can define
a gauge model over a matrix algebra. The standard steps, like writing
down a volume form and a star operation, introducing $\delta = * \, d
\, *$ and defining the Laplace Beltrami operator and the Lie
derivative, can be done as in the commutative case.

\vglue 1cm

\noindent {\bf 2. The Fuzzy Sphere}

\medskip

Let us first give a suitable description of the set of smooth
functions defined over the usual sphere $S^2 \hookrightarrow {\bf R}^3$.
Let $X_i$ denote three commuting coordinates $[X_i ,X_j ] = 0$. The
algebra we are dealing with is given by
$$
\Ac_{\infty} = \{ f(X_1 ,X_2 ,X_3) \mid f\hbox{-analytic}\} /I \, ,
\eqno (2)
$$
where the ideal $I$ consists of functions vanishing at $\build
\sum_{i=1}^{3} \, X_i^2 = R^2$. We are interested in models which are
invariant under rotations of the sphere. The generators of rotations
are given by
$$
L_i = \epsilon_{ijk} \, X_j \,
{1\over i} \, {\partial \over \partial \, X_k} \, ,
\eqno (3)
$$
which obey the $su(2)$ commutation relations. An invariant action
corresponding to a free 2-dimensional scalar field $\phi$ is given by
$$
S \, [\phi] = \sum_{i=1}^3 \, \langle L_i \, \phi \mid L_i \, \phi
\rangle \, , \eqno (4)
$$
where we have introduced the scalar product
$$
\langle \phi \mid \psi \rangle = \int {d^3 \, X \over 2 \pi \, R} \,
\delta ({\vec X}^2 -R^2) \, \phi^+ (X) \, \psi (X) \eqno (5)
$$
within the algebra $\Ac_{\infty}$.

\smallskip

\noindent From $SU(2)$ transformation properties, we deduce that
$X_i$ transforms according to the spin 1 irreducible representation
and higher order products of $X_i$'s transform according to higher
spin representations. Our algebra can therefore be decomposed in the
following way
$$
\Ac_{\infty} = [0] \oplus [1] \oplus \cdots \oplus [j] \oplus \cdots \, ,
\eqno (6)
$$
where $[j]$ means the vector space of the spin $j$ representation.

\smallskip

The {\it truncation} or quantization of $\Ac_{\infty}$ is now {\it
defined} as the family of noncommutative algebras $\Ac_j$ given by
the truncated sum of irreducible spin $j$ representations [3]
$$
\widehat{\Ac}_j = [0] \oplus [1] \oplus \cdots \oplus [j] \, , \eqno (7)
$$
equipped with an associate product and scalar product, which give in
the limit $j\rightarrow \infty$ the standard product in $\Ac_{\infty}$.
In order
to define product we consider the space $\Lc \, \left( {j\over 2} \,
, \, {j\over 2} \right)$ of linear operators from the representation
space for spin $j/2$ to itself. $SU(2)$ acts on $\Lc \, \left(
{j\over 2} \, , \, {j\over 2} \right)$ by the adjoint action. This
representation is reducible and the standard Clebsch-Gordan
decomposition implies that $\Lc \, \left( {j\over 2} \, , \, {j\over 2}
\right) \equiv \Ac_j$. By the standard matrix multiplication in $\Lc
\, \left( {j\over 2} \, , \, {j\over 2} \right)$ we obtain a
noncommutative product in $\Ac_j$. As a scalar product, we take
$$
\langle f \mid g \rangle_j = {1\over j+1} \, {\rm Tr} \, f^+ \, g \ , \ f,g
\in \Ac_j \, . \eqno (8)
$$

We are now able to make more precise in which way the product and
scalar product so defined converge to their commutative limits. There
is a natural chain of embeddings of vector spaces
$$
{\Ac}_0 \hookrightarrow {\Ac}_1 \hookrightarrow {\Ac}_2 \hookrightarrow
\ldots {\Ac}_j \hookrightarrow \ldots {\Ac}_{\infty} \, . \eqno (9)$$
Any normalized element from ${\Ac}_j$ of the form $C_{\ell
m}^{(j)} \, L_-^m \, X_{(j)}^{+\ell}$ is mapped to a normalized
element from $\Ac_k$ given by $C_{\ell m}^{(k)} \, L_-^m \,
X_{(k)}^{+\ell}$. Here $X_{(j)}^{\alpha}$ denote the representatives of
the $su(2)$ generators in the irreducible representation with spin
$j/2$. They obey the relations [4]
$$
[X_{(j)}^m \, , \, X_{(j)}^n ] = i \, {R \, \epsilon_{mnp} \over
\sqrt{{j\over 2} \, \left( {j\over 2} +1 \right)}} \, X_{(j)}^p \, ,
\eqno (10)
$$
and the normalization coefficients $C_{\ell m}^{(j)}$ and $C_{\ell
m}^{(k)}$ are determined through the requirement that the embedding
conserves the norm $\Vert X_{(j)}^i \Vert_j^2 = {R^2 \over 3}$. We
should also note that the defining relation for the surface $S^2$
$$
\sum_{i=1}^3 \, (X_{(j)}^i )^2 = R^2 \eqno (11)
$$
follows from (10) and holds therefore in the noncommutative case, too.

\smallskip

According to (9), elements $f,g$ of $\Ac_j$ can be canonically
considered as elements of $\Ac_k$ with $k \geq j$. Their product in
every $\Ac_k$ can be embedded in $\Ac_{\infty}$ too. If we denote it by
$(fg)_k$ we can prove that

\medskip

\noindent {\ibf Lemma:}
$$
\build \lim_{k\rightarrow \infty}^{} \ (f \, g)_k = f \, g \, , \eqno (12)
$$
{\it where on the rhs. the commutative multiplication in $\Ac_{\infty}$
is meant. As for the proof, we have to show that the coefficients
$C_{\ell m}^{(k)}$ converge to their commutative analog: $\build
\lim_{k\rightarrow \infty}^{} \ C_{\ell m}^{(k)} = C_{\ell m}^{\infty}$, which
can be done after the norm of $\Vert (X_{(k)}^+)^{\ell} \Vert_k^2 =
(C_{\ell 0}^{(k)})^{-2}$ is evaluated.}

\medskip

We may remark that the truncated sphere was introduced also by
Berezin [5], who quantized the symplectic two-form on the ordinary
two-sphere. Hoppe [6] investigated properties of spherical membranes
and introduced a truncation of high frequency excitations, which led
him to the quantum sphere too. Motivated by noncommutative geometry,
Madore has reinvented the Fuzzy sphere [4] and has provided an important
change of the viewpoints: instead of understanding the non-commutative
manifold as the {\it final} product of the deformation quantization he
developed
some basic differential geometry {\it starting} from the notion of the
non-commutative manifold. Field theories on the Fuzzy
sphere were formulated in [2] and [3, 7, 8, 9].

\smallskip

It is interesting to remark, that a different approach, the so called
orbit method using coherent states leads to the same structure [10].
One starts with a group $G$ (in our case $SU(2)$) and a unitary
irreducible representation $T(g)$, $g\in G$ in a Hilbert space $\Hc$.
Let $\vert 0 \rangle \in \Hc$, then $T(g) \vert 0 \rangle$ has
a stability group $H$ (in our case $U(1)$), and the set $\{ T(g) \vert
0 \rangle \}$ parametrizes the homogeneous space $G/H$ (which
becomes in our case $S^2$). These states $\{ \vert x \rangle
\mid x \in G/H \}$ are over complete $\int d\mu (x) \vert x
\rangle \langle x\vert = \un$ and allow to quantize
$f\rightarrow \widehat f = \int f(x) \, \vert x \rangle \,
\langle x\vert \, d\mu (x)$ and to dequantize $f(x) = \langle x \vert
\, \widehat f \, \vert x \rangle$ functions over $G/H$.
Convolution on the group defines the so called star-product. Applying
the star product to the coordinate functions gives a noncommutative
product. In case $T(g)$ is taken to be the spin $j$ representation of
$SU(2)$ the matrix algebra we have discussed before, results.

\smallskip

{}From the last remarks it becomes clear, that our procedure works not
only for $SU(2) /$ $U(1) \simeq S^2$, but more general. We did work out
[10] a few more details for the factor space $SU(1,1) / U(1)$ which
becomes a two-sided hyperboloid. Clearly the appropriate algebra
becomes then infinite dimensional. In addition, the group $SU(1,1)$
has more than one series of unitary representations.

\smallskip

In the following we shall deal only with $S^2$ and finite matrix
approximations. Physics is then reduced to finite degrees of freedom.
The sphere is covered by a finite number of cells. In this way we
introduce a fundamental length into the system, although our main
interest concerns the fact that an ultraviolet cutoff results.

\vglue 1cm

\noindent {\bf 3. Scalar Field Theory}

\medskip

{}From the notions introduced above it is easy to formulate a scalar
field theory. The truncated action may be taken to be
$$
S_N \, [\phi^+ ,\phi] = {1\over N+1} \, {\rm Tr}_N \{ L_i \, \phi^+ \, L_i
\, \phi + {\rm Pol} \, (\phi^+ ,\phi) \} \, ,\quad N=j, \eqno (13)
$$
where ${\rm Pol}(.,.)$ is a positive polynomial.
Expectation values of observables $F(\phi^+ ,\phi)$ are now given by
$$
\langle F \rangle_N = {1\over Z_N} \, \int [d \, \phi^+ \, d \, \phi
]_N \, e^{-S_N [\phi^+ ,\phi]} \, F(\phi^+ ,\phi) \, , \eqno (14)
$$
where the measure means integrating over all $(N+1)^2$ matrix elements.

\smallskip

We therefore obtained an ultraviolet cut-off through the use of the
noncommutative space. But, in addition, the space symmetries, that is
to say, rotations of the sphere, leave the action invariant, as long
as the polynomial in equation (13) is rotational invariant [9]. This
is one of the novel features. We approximated the quantum field by a
finite number of modes, but kept all the symmetries of the model.

\smallskip

We note that reflection positivity is obeyed too and the
Osterwalder-Schrader axioms hold.

\smallskip

Although we are mainly concerned with nonperturbative aspects, the
Feynman rules are of interest too [9]. The free propagation is given
by $\delta_{\ell \ell'} \, \delta_{mm'} \, {1\over \left( \ell +
{1\over 2}\right)^2}$. For a $(\phi^+ \, \phi)^2$ interaction, for
example the four vertex introduces a nonlocality. Locality is obtained
for $N\rightarrow \infty$, but the old difficulties of quantum field
theory show up, too. The tadpole graph, for example, will diverge like
${\rm ln}\, N$.

\smallskip

We developped actually a second way to quantize fields defined over
$S^2$ [11]. In addition to results obtained before, this procedure
will allow us to deal with nontrivial topological configurations, or
in other words with sections of $U(1)$-bundles over $S^2$ and their
matrix approximations. Classical topological nontrivial
configurations on $S^2 \hookrightarrow {\bf R}^3$ are classified with the
help of the Hopf fibration. We start from ${\bf C}^2 \ni (\chi_1 ,\chi_2)$
and restrict the two complex coordinates $\chi_{\alpha}$ to lie on
$S^3$ of radius $\sqrt R : \vert \chi_1 \vert^2 + \vert \chi_2 \vert^2
=R$. Next we introduce the mapping $\chi_{\alpha} \rightarrow X_i =
\chi^+ \, \sigma_i \, \chi \in {\bf R}^3$. The restriction of
$\chi_{\alpha}$ to the three-sphere $S^3$ of radius $\sqrt R$ implies
that $X_i$ lies on $S^2 : \, \build \sum_{i}^{} \, (X_i)^2 = R^2$.
Since $X_i$'s do not change under the transformation
$$
\chi \rightarrow e^{{i\over 2} \psi} \, \chi \quad , \quad \chi^+ \rightarrow
e^{-{i\over 2} \psi} \, \chi^+ \, , \eqno (15)
$$
we see that the fiber is $U(1)$.

\smallskip

As $\Ac_k$, $k\in {1\over 2} \, Z$, we denote the linear space of
functions in ${\bf C}^2$ (or $S^3$ after the restriction) of the form
$$
\displaylinesno{
\Ac_k = &(16) \cr
\{ f(\chi^+ ,\chi) = \sum \, a_{m_1 m_2 n_1 n_2} \,
(\chi_1^+)^{m_1} \, (\chi_2^+)^{m_2} \, (\chi_1)^{n_1} \,
(\chi_2)^{n_2} \mid 2k = m_1 + m_2 - n_1 - n_2 \} \, . \cr
}
$$
Under the transformation (15) $f\in \Ac_k$ goes into $e^{-ik\psi} \,
f$. They are eigenfunctions of the operator $K_0 = {1\over 2}
(\chi_{\alpha}^+ \, \partial_{\chi_{\alpha}^+} - \chi_{\alpha} \,
\partial_{\chi_{\alpha}})$ with eigenvalue $k$. There is an involutive
gradation with $\Ac_k^+ = \Ac_{-k}$ and $\Ac_k \, \Ac_e \subset
\Ac_{k+e}$. $\Ac_0$ is the algebra of polynomials in the variables
$X_i$ and  $\Ac_k$ are $\Ac_0$-modules.

\smallskip

The generators of rotations in $S^2$ can now be expressed in terms of
the new variables as
$$
L_i = {i\over 2} \, (\chi_{\alpha}^+ \, \sigma_{\alpha\beta}^{*i} \,
\partial_{\chi_{\beta}^+} - \chi_{\alpha} \, \sigma_{\alpha \beta}^i \,
\partial_{\chi_{\beta}}) \, . \eqno (17)
$$
They satisfy the $su(2)$ algebra in $\Ac_k$ and leave invariant the
function defining the sphere: $L_i \build \sum_{k}^{} \, (\chi^+ \,
\sigma_k \, \chi)^2 =0$.

\smallskip

We shall actually work with two independent derivatives
(corresponding to the \break restriction to the sphere) defined by
$$
K_+ \, f = i \, \epsilon_{\alpha \beta} \, \chi_{\alpha}^+ \,
\partial_{\chi_{\beta}} \, f \quad , \quad K_- \, f = i \, \epsilon_{\alpha
\beta} \, \partial_{\chi_{\alpha}^+} (f) \, \chi_{\beta} \, . \eqno
(18)
$$
The operators $K_+ \, , \, K_- \, , \, K_0$ satisfy $su(2)$ algebra
relations. The following relation holds between Casimir operators of the
representations (17) and (18)
$$
\sum_i \, L_i^2 = K_0^2 + {1\over 2} \, (K_+ \, K_- + K_- \, K_+)
\, . \eqno (19)
$$
The action of a complex field $\phi \in \Ac_k$ with topological
charge $2k$ is now given in terms of the scalar product introduced in
(5) as
$$
S_k \, [\phi] = \langle \phi^+ \, \vert K_+ \, K_- + K_- \, K_+ \vert \,
\phi \rangle + \langle 1 \vert V (\phi^+ ,\phi)\rangle \, .
\eqno (20)
$$
We describe next the noncommutative version of the above steps using
the Jordan-Schwin\-ger realization of $su(2)$ [11]. This way we obtain
also a noncommutative version of the Hopf fibration. We replace
coordinates $\chi_{\alpha}^+$ and $\chi_{\alpha}$ by the following
combinations of creation and
annihilation operators $A_{\alpha}^+$ and $A_{\alpha}$
$$
\chi_{\alpha}^+ \rightarrow {1\over\sqrt{\hat N}} \, A_{\alpha}^+
\quad , \quad \chi_{\alpha} \rightarrow  \,
A_{\alpha} \, {1\over\sqrt{\hat N}}, \,
\hat N= A_1^+ A_1 +A_2^+ A_2.\eqno (21)
$$
Here $A_{\alpha}^+,A_{\alpha}$ obey the canonical commutation relations:
$$
[A_{\alpha} \, , \, A_{\beta}^+] = \delta_{\alpha \beta} \quad , \quad
[A_{\alpha} \, , \, A_{\beta}] = [A_{\alpha}^+ \, , \, A_{\beta}^+] = 0
\, . \eqno (22)
$$
 We represent
$A_{\alpha}^+$ and $A_{\alpha}$ in the bosonic Fock space $\Fc$ spanned
by vectors
$$
\vert n_1 ,n_2 \rangle = {1\over \sqrt{n_1! \, n_2!}} \ (A_1^+)^{n_1}
\, (A_2^+)^{n_2} \vert 0 \rangle \, , \eqno (23)
$$
where the vacuum is defined by $A_1 \vert 0 \rangle = A_2 \vert 0
\rangle =0$. The operators
$$
X_k = {1\over 2} \, A_{\alpha}^+ \, \sigma_{\alpha \beta}^k \,
A_{\beta} \eqno (24)
$$
fulfil the $su(2)$ algebra relations in $\Fc$. In what follows, we
need restrictions of $\Fc$ to the $(N+1)$-dimensional subspaces
$\Fc_N \subset \Fc$,
$$
\Fc_N = \{ \vert n_1 ,n_2 \rangle \mid n_1 + n_2 =N \} \quad , \quad
N=0,1,2,\ldots \eqno (25)
$$
on which the $su(2)$ algebra is represented irreducibly. The Casimir
operator $C = X_3^2 + {1\over 2} \, (X_+ \, X_- + X_- \, X_+)$ takes
the value ${N \over 2} \, \left( {N\over 2} +1 \right)$ on $\Fc_N$.
Appropriately scaled we arrive again at the noncommutative ``sphere''.

\smallskip

As $\widehat{\Ac}_{MN}$ we denote the space of linear mappings from
$\Fc_N$ to $\Fc_M$ spanned by monomials $(A_1^+)^{n_1} \,
(A_2^+)^{n_2} \, (A_1)^{m_1} \, (A_2)^{m_2}$ with $n_1 + n_2 \leq M$,
$n_1 + n_2 \leq N$ and $2k =M-N=n_1 +n_2 -m_1 -m_2$. We obviously
have $\widehat{\Ac}_{MN}^+ = \widehat{\Ac}_{NM}$, $\widehat{\Ac}_{LM} \,
\widehat{\Ac}_{MN} \subset \widehat{\Ac}_{LN}$. Operators from
$\widehat{\Ac}_{MN}$ maps $\Fc_N$ to $\Fc_M$ and can be represented by
$(N+1)$ times $(M+1)$ dimensional matrices. There is an antilinear
isomorphism between $\widehat{\Ac}_{MN}$ and $\widehat{\Ac}_{NM}$
given by hermitian conjugation of these matrices. In particular
$\widehat{\Ac}_{NN}$ is the $(N+1)^2$-dimensional algebra of $(N+1)
\times (N+1)$ square matrices. Obviously, $\widehat{\Ac}_{MN}$ is a
left $\widehat{\Ac}_{MM}$ -- and a right $\widehat{\Ac}_{NN}$ module.

\smallskip

Generators $L^j$ act in $\widehat{\Ac}_{MN}$ as
$$
L^j \, f = X_{(M)}^j \, f - f \, X_{(N)}^j \, , \eqno (26)
$$
where $X_{(N)}^j$ denotes the representation of the operator $X^j$ in
$\Fc_N$. This $su(2)$ representation is reducible and equivalent to
the direct product of two $su(2)$ representations
$$
\left[ {M \over 2} \right] \otimes \left[ {N \over 2} \right] = [\vert
k \vert ] \oplus [\vert k \vert +1] \oplus \ldots \oplus \left[ {M+N
\over 2} \right] \, , \eqno (27)
$$
where $2k = M-N$. This means that any operator $f\in \widehat{\Ac}_{MN}$
can be expanded into a base of operators transforming according to
(27).

\smallskip

We remark, that the description of topologically nontrivial field
configurations with $2k \ne 0$ in the noncommutative setting needs
two algebras $\widehat{\Ac}_{MM}$ and $\widehat{\Ac}_{NN}$ with $2k = M-N$. A
noncommutative analog of a complex scalar field $\phi$ with fixed
winding number $k$ we identify as element of $\widehat{\Ac}_{MN}$. For the
action we take
$$
S_{MN} \, [\phi] = {2\over M+N+2} \, {\rm Tr}_N \, (\phi^+ \, (K_+ \, K_- +
K_- \, K_+ ) \, \phi + V(\phi^+ ,\phi)) \, , \eqno (28)
$$
where the operators $K_{\pm}$ are defined by
$$
K_+ \, \phi = i \, \epsilon_{\alpha \beta} \, A_{\beta}^+ \, [\phi
,A_{\alpha}^+] \quad , \quad K_- \, \phi = i \, \epsilon_{\alpha \beta} \,
[A_{\alpha} ,\phi] \, A_{\beta} \eqno (29)
$$
and $2K_0 = {\rm ad} \, (A_{\alpha}^+ \, A_{\alpha})$ takes on the
constant value $2k$. The order of operators in (29) is essential, so
that they act like in the commutative case. Moreover, putting
 $2J = M+N$ and keeping $k$ fixed, we
recover the commutative action in the limit $J \rightarrow \infty$. We
conclude, that we obtain this way a fully $su(2)$ symmetric, that
means rotation symmetric model, which is described by a finite number
of modes. In a certain sense, we work on a noncommutative finite
``lattice''.

\vglue 1cm

\noindent {\bf 4. The Spinor Field}

\medskip

The spinor bundle over $S_2$ is a standard part of any textbook of
quantum mechanics. Also the spectrum of the Dirac operator and all
eigenfunctions (the spinorial harmonics) are known, and we are
therefore very brief. We may start from the spinor bundle over
${\bf R}^3$ with sections being two component spinorial wave functions
$(\psi_+ ,\psi_-)$. The action of $su(2)$ is described by the
generators $J_i = L_i + {\sigma_i \over 2}$. View ${\bf R}^3$ as ${\bf R}_+
\times S^2$. A simple exercise allows to express the flat Dirac
operator in three dimensions in terms of spherical coordinates. This
way we obtain the Dirac operator corresponding to the round metric on
$S^2$ in terms of the $su(2)$ generators as follows [3]
$$
D = {1\over R} \, (\vec{\sigma} \cdot \vec L +1) \, , \eqno (30)
$$
$D$ is self-adjoint with respect to the scalar product
$$
\langle \psi \mid \varphi \rangle = \int \, {d^3 \, X \over 2\pi R} \,
\delta (\vec{X}^2 -R^2) \, (\psi_+^+ \, \varphi_+ + \psi_-^+ \,
\varphi_-) \, , \eqno (31)
$$
where $\psi = (\psi_+ ,\psi_-)$ and $\varphi = (\varphi_+ , \varphi_-)$
belong to the spinor bundle $S_{\Ac_{\infty}}$. From the point of view
of transformation properties under $su(2)$, $S_{\Ac_{\infty}}$
transforms according to the tensor product of the spin $1/2$
representation $[1/2]$ times representations of $\Ac_{\infty}$:
$$
S_{\Ac_{\infty}} = 2 \cdot \left( \left[ {1\over 2} \right] \oplus \left[
{3\over 2} \right] \oplus \left[ {5\over 2} \right] \oplus \cdots \right)
\, . \eqno (32)
$$
All half-integer spin representations occur exactly twice. The
doubling corresponds to left and right chiral spinor bundles. It is
equally easy to work out the Dirac operator $D_k$ on spinors with
components $\psi_{\alpha}$, which are sections belonging to the set
$\Ac_k$ (16). They become
$$
D_k = {1\over R} \, (\vec{\sigma}  \cdot \vec J +1+k \, \vec{\sigma}
\cdot \vec x ) \, , \eqno (33)
$$
where $\vec{\sigma} \cdot \vec x / R = \Gamma$ equals the chirality
operator. The winding number $k$ equals the monopole charge. The
spectrum of $D_k$ from (33) is given by
$$
E_{k,j} = \pm \sqrt{\left( j + {1\over 2}\right)^2 -k^2} \, ,
\eqno (34)
$$
where $j=\vert k \vert + 1/2$, $\vert k \vert + 3/2 , \ldots$
corresponds to the non-zero-eigenvalue modes. $j = \vert k \vert
-{1\over 2}$ gives zero modes and is admissible only for $k \ne 0$.
For $k>0$ negative chirality zero modes exist, while for $k<0$
positive chirality zero modes appear.

\smallskip

We next look for a noncommutative analog of the spinor bundle. A
counting of degrees of freedom of vector spaces entering in (32)
shows that the naive approach of considering the elements of
${\widehat{\Ac}}_{MN}$ as the spinor components does not work. The
noncommutative
sphere described in chapter two emerged from the quantization of the
algebra of scalar functions on the ordinary sphere. It was therefore
natural to treat bosons and fermions simultaneously and to study the
supersymmetric extension of the above scheme. We found that a
consistant truncation (deformation quantization) of the ring of
scalar superfields is possible. The fermions are then identified as
components of these superfields. In the topologically non-trivial case,
 we have obtained the quantization [11] by deforming the
$N=1$ superextension of the Hopf fibration described in chapter three.

\smallskip

As the Hopf superfibration [11] we denote the mapping from
${\bf C}^{2,1}$ to ${\bf R}^{3,2}$ given by
$$
\xi = \left( \matrix{ \chi_1 \cr \chi_2 \cr a \cr} \right)
\leftrightarrow (X_i ; \theta_+ , \theta_-) \quad , \quad i=1,2,3 \eqno
(35)
$$
$$
X_i = \xi^+ \, R_i \, \xi \qquad , \qquad \theta_{\pm} = \xi^+ \,
F_{\pm} \, \xi
$$
where $\chi_i$ are two complex even parameters, while $a$ is a
complex odd parameter. $R_3 , R_{\pm} = R_1 \pm iR_2$ and $F_{\pm}$
are $3\times 3$ matrices given as
$$
F_+ = {1\over \sqrt 2} \ \left( \matrix{ 0 &0 &-1 \cr 0 &0 &0 \cr 0
&-1 &0 \cr} \right) \quad , \quad F_- = {1\over \sqrt 2} \ \left(
\matrix{ 0 &0 &0 \cr 0 &0 &-1 \cr 1 &0 &0 \cr} \right)
$$
$$\eqno (36)$$
$$
R_+ = {1\over 2} \ \left( \matrix{ 0 &1 &0 \cr 0 &0 &0 \cr 0
&0 &0 \cr} \right) \ , \ R_- = {1\over 2} \ \left( \matrix{ 0 &0
&0 \cr 1 &0 &0 \cr 0 &0 &0 \cr} \right) \ , \ R_3 =
\left( \matrix{ 1 &0 &0 \cr 0 &-1 &0 \cr 0 &0 &0 \cr} \right) \, .
$$
They generate the ${\rm osp} \, (2,1)$ superalgebra [3]. $F_+$ and
$F_-$ are the odd generators. Their brackets gives the three even ones:
$$
\{ F_+ , F_- \} = -R_3 \quad , \quad \{ F_{\pm} , F_{\pm} \} = \pm
R_{\pm} \, . \eqno (37)
$$
$R_{\pm}$ and $R_-$ obey $su(2)$ algebra relations among themselves
and even and odd relations
$$
[R_3 , F_{\pm}] = \pm {1\over 2} \, F_{\pm} \quad , \quad [R_{\pm}
,F_{\pm}] = 0 \quad , \quad [R_{\pm} , F_{\mp} ] = F_{\pm} \eqno (38)
$$
show that $F_{\pm}$ transforms like a $su(2)$ spinor and the
superalgebra closes.

\smallskip

A superfunction is now defined as a linear combination of monomials
in $\chi_{\alpha}^+$, $\chi_{\alpha}$ and $a^+ ,a$. To such a monomial
we assign a topological charge $2k$ and define the set of superfunction
$$
\displaylinesno{
(s \, \Ac_{\infty})_k = \Bigl\{ f = \sum a_{m_1 m_2 n_1 n_2 \mu \nu} \,
(\chi_1^+)^{m_1} \, (\chi_2^+)^{m_2} \, (\chi_1)^{n_1} \,
(\chi_2)^{n_2} \, a^{+\mu} \, a^{\nu} \mid \cr
2k = m_1 +m_2 -n_1 -n_2 +\mu -\nu \Bigl\} &( 39) \cr
}
$$
$\phi \in (s \, \Ac_{\infty})_k$ can be expanded as
$$
\phi = \phi_0 (\chi^+ ,\chi) + f(\chi^+ ,\chi) \, a + g(\chi^+ ,\chi)
\, a^+ + F(\chi^+ ,\chi) \, a^+ \, a \, , \eqno (40)
$$
where $\phi_0 ,F \in (\Ac_{\infty})_k$, $f\in (\Ac_{\infty})_{k+1/2}$ and
$g\in
(\Ac_{\infty})_{k-1/2}$. Much as before there exist an involutive
gradation. $(s \, \Ac_{\infty})_0$ is a superalgebra with respect to the
supercommutative product of parameters $\xi^+$ and $\zeta$, $(s \,
\Ac_{\infty})_k$ are $(s \, \Ac_{\infty})_0$-bimodules.

\smallskip

The differential operators generating the ${\rm osp} \, (2,1)$ algebra
in $(s \, \Ac_{\infty})_k$ are given by
$$
\displaylinesno{
J_k = {i\over 2} \, (\xi_{\alpha}^+ \, R_{\alpha \beta}^k \,
\partial_{\xi_{\beta}^+} - \xi_{\alpha} \, R_{\alpha \beta}^{*k} \,
\partial_{\xi_{\beta}}) \cr &(41) \cr
V_{\pm} = {i \over 2} \, (\xi_{\alpha}^+ \, F_{\alpha \beta}^{\pm} \,
\partial_{\xi_{\beta}^+} + \xi_{\alpha} \, (F_{\alpha \beta}^{\pm})^*
\, \partial_{\xi_{\beta}}) \, . \cr
}
$$
We deduce by explicite calculation, that $C(X,\theta) = \, \build
\sum_{i}^{} \, X_i^2 + {1\over 2} \, (\theta_+ \, \theta_- - \theta_-
\, \theta_+)$ is an invariant superfunction from $(s \, \Ac_{\infty})_0$:
$$
J_i \, C (X,\theta) = V_{\pm} \, C(X,\theta) = 0 \, .
$$
We can therefore define the supersphere in ${\bf R}^{3,2}$ by the
condition that $C(X,\theta) = R^2$. Note that  the
superfunctions
are understood to be restricted to the supersphere.
The Berezin integral times the standard measure on the sphere allows
to introduce an inner product over the supersphere.

\smallskip

Before turning to the non commutative truncation, we observe that
irreducible representations of ${\rm osp} \, (2,1)$ consist of pairs
of $su(2)$ irreducible representations $[ \widetilde j ] = [j] \oplus
\left[ j-{1\over 2}\right]$, where $j$ is either an integer or a
half-integer. The generators $X_i , \theta_{\pm} \in (s \, \Ac_{\infty})_0$
form
a superspin $1$ irreducible representation of the ${\rm osp} \, (2,1)$
algebra under the action of the vector fields (41) and higher powers
of $X_i$ and $\theta_{\pm}$ can be arranged into irreducible
supermultiplets of higher superspins. From the point of view of
$su(2)$ representations, the algebra $(s \,
\Ac_{\infty})_0$ of superfields consists of two
copies of $\Ac_{\infty}$ and the spinor bundle $\left[ {1\over 2}
\right] \otimes \Ac_{\infty}$. This gives the full decomposition of $(s
\, \Ac_{\infty})_0$ into irreducible representations of ${\rm osp} \,
(2,1)$ as an infinite direct sum
$$
(s \, \Ac_{\infty})_0 = \widetilde{[0]} \oplus \widetilde{\left[ {1\over 2}
\right]} \oplus \widetilde{[1]} \oplus \widetilde{\left[ {3\over 2}
\right]} \oplus \cdots \, . \eqno (42) $$
It is instructive to insert into (42) the spin representations and
count the degrees of freedom. Whenever you truncate they sum up to a
complete square.

\smallskip

Thus, we proceed as in chapter two and define the truncations of $(s
\, \Ac_{\infty})_0$ as the family of noncommutative superspheres $(s \,
\Ac_j)_0$, being given by the sum of irreducible representations of
${\rm osp} \, (2,1)$
$$
(s \, \Ac_j)_0 = \widetilde{[0]} \oplus \widetilde{\left[ {1\over 2}
\right]} \oplus \widetilde{[1]} \oplus \widetilde{\left[ {3\over 2}
\right]} \oplus \dots \oplus \widetilde{[j]} \quad , \quad j \in Z \eqno (43)
$$
together with an associative product and an inner product such that
in the limit $j\rightarrow \infty$ the standard product of
$(s \, \Ac_{\infty})_0$
is obtained. We can represent the algebra $(s \, \Ac_j)_0$ in the
Fock space $(s \, \Fc)$ of bosonic and fermionic degrees of freedom.
We replace
$$
(\chi_{\alpha}^+ , \chi_{\alpha}) \rightarrow
\left( {1\over \sqrt{s\hat{N}}} \, A_{\alpha}^+ ,
A_{\alpha} {1\over \sqrt{s\hat{N}}}\right), ~ (a^+ ,a) \rightarrow
\left({1\over \sqrt{s\hat{N}}} \, b^+ ,  b {1\over \sqrt{s\hat{N}}}\right),~
 s\hat{N}= \hat N + b^+ b ,\eqno (44)
$$
where $b^+ ,b$ are fermionic creation and annihilation operators
and $\hat N$ was defined in (21). $(s
\, \Fc )$ is spanned by vectors
$$
\vert n_1 ,n_2 ; \nu \rangle = {1\over \sqrt{n_1 ! \, n_2 !}} \,
(A_1^+)^{n_1} \, (A_2^+)^{n_2} \, (b^+)^{\nu} \, \vert 0 \rangle \eqno
(45)
$$
with $n_i \geq 0$, $\nu =0,1$, and the supervacuum is defined by
$A_{\alpha} \, \vert 0 \rangle = b \, \vert 0 \rangle = 0$. The
operators  $$
\widehat{X}_j = {1\over 2} \, A_{\alpha}^+ \, \sigma_{\alpha \beta}^j
\, A_{\beta} \quad , \quad \widehat{\theta}_+ = {-1 \over \sqrt 2} \,
(A_1^+ \, b + A_2 \, b^+) \quad , \quad \widehat{\theta}_- = {1\over
\sqrt 2} \, (-A_2^+ \, b + A_1 \, b^+) \eqno (46)
$$
satisfy in $(s \, \Fc)$ the ${\rm osp} \, (2,1)$ superalgebra
commutation relations (37) and (38). The ``superadjoint'' action of
the form $$
\widehat{J}_k \, \phi = [\widehat{X}_k ,\phi] \quad , \quad \widehat{V}_{\pm} =
[\widehat{\theta}_{\pm} ,\phi]_g \, , \eqno (47)
$$
where $[\cdot , \cdot ]_g$ denotes the graded commutator, defines a
reducible representation of ${\rm osp} \, (2,1)$. As
$$
(s \, \Fc)_N = \{ \vert n_1 ,n_2 ; \nu \rangle \in (s \, \Fc) \mid n_1
+n_2 +\nu =N \} \eqno (48)
$$
we denote a $(2N+1)$-dimensional subspace, which may be decomposed
into a subspace with $N$ bosons and no fermions and a subspace with
$(N-1)$ bosons and one fermion. $(s \, \Fc)_N$ is the representation
space of an irreducible representation of ${\rm osp} \, (2,1)$ where
the Casimir operator
$$
C=\widehat X_3^2 + {1\over 2} \, \{ \widehat X_+ , \widehat X_- \} +
{1\over 2} \, [\widehat {\theta}_+ ,
\widehat{\theta}_-] \eqno (49)
$$
takes on the value $N(N+1)/4$. As $(s \, \widehat{\Ac})_{MN}$ we denote the
space of linear mappings from $(s \, \Fc)_N$ to $(s \, \Fc)_M$
spanned by operator monomials of the form (39) with $m_1 + m_2 + \mu
\leq M$, $n_1 +n_2 +\nu \leq N$ and $m_1 +m_2 +\mu -n_1 -n_2 -\nu =
M-N$. These operators are represented by $(2N+1) \times (2M+1)$
matrices. In $(s \, \widehat{\Ac})_{MN}$ an inner product is defined by
the supertrace $\langle \phi /\psi \rangle_{MN} =  \, s \,
{\rm Tr}\, (\phi^+ \, \psi)$ in the space of linear operators
from $(s \, \Fc)_N$ to  $(s \, \Fc)_N$.

\smallskip

The spinor field we identify in the supercommutative case as the odd
part of the superfield $\phi$ of equation (40):
$$
\psi = f \, (\chi^+ , \chi) \, a + g \, (\chi^+ ,\chi) \, a^+ \, ,
\eqno (50)
$$
$f$ and $g$ correspond to chirality eigenmodes.

\smallskip

$\Gamma$ maps $\psi$ into $\Gamma \, \psi = f \, a - g \, a^+$. The
Dirac operator maps from $(s \, \Ac)_k$ to $(s \, \Ac)_k$ as
$$
D \, \psi = (K_+ \, g) \, a + (K_- \, f) \, a^+ \, , \eqno (51)
$$
and anticommutes with $\Gamma : \{ D ,\Gamma \} =0$.

Next we return to
the truncated situation. As $\widehat{S}_k$, $k\in {1\over 2} \, Z$,
we denote the set of odd elements from $(s \, \widehat{\Ac})_k$, with
$f\in \widehat{\Ac}_{k+1/2}$ and $g \in \widehat{\Ac}_{k-1/2}$,
where $(s \, \widehat{\Ac})_k$ are $(s \, \widehat{\Ac})_0$-modules
generated by the operators (44) before the restriction to a particular
$(s \, \Fc)_N$. Note that
$\widehat{S}_k$ are $\widetilde{\Ac}_0$-bimodules but not $(s \,
\widehat{\Ac})_0$-bimodules. Equation (50) with
non-commutative $\hat{\chi}^+ , \hat{\chi} ,\hat a$ given by (44)
again induces a decomposition of the spinor space
$\widehat{S}_k$ into the direct sum of positive and negative chirality
contribution. The noncommutative analog of the chirality operator
$\widehat{\Gamma}$ acts as
$$
\widehat{\Gamma} \, \psi = -[b^+ \, b,\psi] \eqno (52)
$$
and the free Dirac operator on $\widehat{S}_k$ becomes
$$ \hat D \, \psi = (K_+ \, g) \, \hat a + (K_- \, f) \, \hat{a}^+ \, ,
\eqno (53)
$$
with $\{ \widehat{\Gamma} , \widehat D \} =0$.

\smallskip

As $\widehat{S}_{MN}$ we denote odd mappings from $(s \, \widehat{\Ac})_{MN}$
which are of the form
$$
\psi = f(\hat{\chi}^+ ,\hat{\chi}) \, \hat a +
g \, (\hat{\chi}^+ ,\hat{\chi}) \, \hat{a}^+ \eqno (54)
$$
where $f \in \widehat{\Ac}_{M,N-1}$ and $g\in \widehat{\Ac}_{M-1,N}$. This
means that $f$ and $g$ can be expanded in terms of operators
transforming according to the representations
$$
\displaylinesno{
f : \left[ {M\over 2} \right] \otimes \left[ {N-1\over 2} \right] =
\left[ k+ {1\over 2} \right] \oplus \cdots \oplus \left[ J -{1\over 2}
\right] \cr
&(55) \cr
g : \left[ {M-1\over 2} \right] \otimes \left[ {N\over 2} \right] =
\left[ k- {1\over 2} \right] \oplus \cdots \oplus \left[ J -{1\over 2}
\right] \, , \cr
}
$$
for $k = {1\over 2} \, (M-N) \ne 0$ and $J = {1\over 2} \, (M+N)$. In
terms of eigen-operators of ${\vec J}^2$ and $J_3$ to eigenvalues $j$
and $m$, $f$ can be expanded in operators with $j=\left\vert
k+{1\over 2}\right\vert, \ldots J-{1\over 2}$, $\vert m \vert \leq
j$, and $g$ into operators with $j = \left\vert k-{1\over
2}\right\vert, \ldots J-{1\over 2}$, $\vert m \vert \leq j$. The
admissible values for $j$ are: $j=\vert k \vert -{1\over 2}$, $\vert
k \vert + {1\over 2}, \ldots ,J-{1\over 2}$, but the first value
$j=\vert k \vert -{1\over 2}$ can occur only for $k \ne 0$. For $k>0$
zero modes occur with negative chirality $(f=0 ,g\ne 0)$, for $k<0$
vice versa they occur with positive chirality $(f\ne 0, g=0)$. The
number of zero modes of the Dirac operator is always $\vert
M-N\vert$. We note that the spectrum of $\widehat D$ on $\widehat{S}_{MN}$ is
identical to that of $D_k$ given by (33), except that it is
cut off at $j={M+N-1 \over 2}$.

\smallskip

The action of a selfinteracting spinor field $\psi \in \widehat{S}_{MN}$
with fixed winding number is defined by
$$
S_{MN} \, [\psi^+ ,\psi] =  s \, {\rm Tr}_N \, (\psi^+ \,
D \, \psi + W (\psi^+ ,\psi)) \, . \eqno (56)
$$
This action is invariant with respect to the following symmetries:

\smallskip

(i) rotations of the sphere,

\smallskip

(ii) chiral transformations $\psi \rightarrow e^{i\alpha
\widehat{\Gamma}} \, \psi$, $\psi^+ \rightarrow \psi^+ \, e^{i\alpha
\widehat{\Gamma}}$,

\smallskip

\noindent as long as the interaction
term in (56) is invariant. Expanding $\psi$ in terms of
eigen-operators of $\widehat D$, a functional integral is given by
integrating over the independent Grassmann expansion coefficients.
Chiral transformations leave invariant this measure $[D \, \psi^+ \,
D \, \psi]_{MN}$ expect for the zero mode contributions. For $k\ne
0$, chiral symmetry is violated on the quantum level and
$$
[D \, \psi^+ \, D \, \psi]_{MN} \rightarrow e^{-i\alpha k} \, [D \,
\psi^+ \, D \, \psi]_{MN} \eqno (57)
$$
under chiral transformations.

\smallskip

Our main tool was to provide a cut-off procedure for simple models,
which respects the symmetries one started with. The supersymmetry
approach allows to describe chiral spinors. The description in terms
of rectangular matrices allows to approximate also topological
nontrivial configurations. A study of coupling to gauge fields has
been done too [8].

\smallskip

Many questions remain, especially: Study of the limit $N\rightarrow \infty$;
different manifolds; more dimensional situations. There may be also an
interesting connection of our formalism with the description of the bound
states of strings and $p$-branes [12] where the space-time coordinates
also enter in the formalism as non-commuting matrices.

\vglue 0.5cm

{\bf Acknowledgement.} In the course of our work we enjoyed
discussions with A. Alekseev, L. \'Alvarez-Gaum\'e, M. Bauer, A. Connes,
V. \v Cern\'y, T. Damour, M. Fecko, J. Fr\"ohlich, J. Ft\'a\v cnik, K.
Gaw\c edzki, J. Hoppe, B. Jur\v co, E. Kiritsis, C. Kounnas, J. Madore,
R. Stora and D. Sullivan.

\smallskip

H.G. thanks the IHES for a pleasant stay, which allowed him to collect
the material.


\bigskip

{\bf References.}

\medskip

\item{[1]} A. Connes; {\it Noncommutative Geometry}, Academic Press
London, 1994.

\smallskip

\item{[2]} J. Madore; {\it Noncommutative Geometry and Applications},
Cambridge University Press, 1995.

\smallskip

\item{[3]} H. Grosse, C. Klim\v c\'\i k and P. Pre\v snajder; {\it Field
Theory on a Supersymmetric Lattice}, hep-th/9507074.

\smallskip

\item{[4]} J. Madore; {\it Math. Phys.} {\bf 32} (1991), 332 and {\it
Class. Quant. Grav.} {\bf 9} (1992), 69.

\smallskip

\item{[5]} F.A. Berezin; {\it Comm. Math. Phys.} {\bf 40} (1975), 153.

\smallskip

\item{[6]} J. Hoppe; {\it Elem. Part. Res. J.} (Kyoto) {\bf 80}
(1989), 145.

\smallskip

\item{[7]} H. Grosse and J. Madore; {\it Phys. Lett.} {\bf B 283}
(1992), 218.

\smallskip

\item{[8]} H. Grosse, C. Klim\v c\'\i k and P. Pre\v snajder; {\it Finite
Gauge Model on Truncated Sphere}, to be published in the Proc. of
Schladming School, 1995.

\smallskip

\item{[9]} H. Grosse, C. Klim\v c\'\i k and P. Pre\v snajder; {\it
Towards Finite Quantum Field Theory in Non-Commutative Geometry},
hep-th/9505175.

\smallskip

\item{[10]} H. Grosse and P. Pre\v snajder; {\it Lett. Math. Phys.}
{\bf 28} (1993), 239.

\smallskip

\item{[11]} H. Grosse, C. Klim\v c\'\i k and P. Pre\v snajder; {\it
Topological Nontrivial Field Configurations in Noncommutative
Geometry}, hep-th/9510083.

\smallskip

\item{[12]} E. Witten; {\it Bound States of Strings and $p$-branes},
hep-th/9510135.

\bye